\documentclass[twoside,twocolumn,9pt]{article}
\usepackage{extsizes}

\usepackage[super,sort&compress,comma]{natbib}
\usepackage[version=4]{mhchem}
\usepackage[left=1.5cm, right=1.5cm, top=1.785cm, bottom=2.0cm]{geometry}
\usepackage{balance}
\usepackage{mathptmx}
\usepackage{sectsty}
\usepackage{graphicx} 
\usepackage{lastpage}
\usepackage[format=plain,justification=justified,singlelinecheck=false,font={stretch=1.125,small,sf},labelfont=bf,labelsep=space]{caption}
\usepackage{float}
\usepackage{fancyhdr}
\usepackage{fnpos}
\usepackage[english]{babel}
\addto{\captionsenglish}{%
  
}
\usepackage{array}
\usepackage{droidsans}
\usepackage{charter}
\usepackage[T1]{fontenc}
\usepackage[usenames,dvipsnames]{xcolor}
\usepackage{setspace}
\usepackage[compact]{titlesec}
\usepackage{hyperref}
\usepackage{pdfpages}
\usepackage{multirow}

\usepackage{epstopdf}

\definecolor{cream}{RGB}{222,217,201}

\usepackage{booktabs}       
\usepackage{subcaption}
\usepackage[list-units=single,detect-all]{siunitx}
\DeclareSIUnit{\calorie}{cal}
\DeclareSIUnit{\kcal}{\kilo\calorie\per\mol}
\DeclareSIUnit{\angstrom}{\text {Å}}
\DeclareSIUnit{\kjoule}{\kJ\per\mol}

\begin{document}

\makeFNbottom
\makeatletter
\renewcommand\LARGE{\@setfontsize\LARGE{15pt}{17}}
\renewcommand\Large{\@setfontsize\Large{12pt}{14}}
\renewcommand\large{\@setfontsize\large{10pt}{12}}
\renewcommand\footnotesize{\@setfontsize\footnotesize{7pt}{10}}
\makeatother

\renewcommand{\thefootnote}{\fnsymbol{footnote}}
\renewcommand\footnoterule{\vspace*{1pt}%
\color{cream}\hrule width 3.5in height 0.4pt \color{black}\vspace*{5pt}} 
\setcounter{secnumdepth}{5}

\makeatletter 
\renewcommand\@biblabel[1]{#1}            
\renewcommand\@makefntext[1]%
{\noindent\makebox[0pt][r]{\@thefnmark\,}#1}
\makeatother 
\renewcommand{\figurename}{\small{Fig.}~}
\sectionfont{\sffamily\Large}
\subsectionfont{\normalsize}
\subsubsectionfont{\bf}
\setstretch{1.125} 
\setlength{\skip\footins}{0.8cm}
\setlength{\footnotesep}{0.25cm}
\setlength{\jot}{10pt}
\titlespacing*{\section}{0pt}{4pt}{4pt}
\titlespacing*{\subsection}{0pt}{15pt}{1pt}

\newcommand{\etal}{\textit{et al.}}

\twocolumn[
  \begin{@twocolumnfalse}

\begin{center}
\LARGE{\textbf{Excited states of polonium(IV): Electron correlation and spin-orbit coupling in the \ce{Po^{4+}} free ion and in the bare and solvated \ce{[PoCl5]-} and \ce{[PoCl6]^{2-}} complexes$^\dag$}}
\end{center}

\vspace{0.3cm}

\begin{center}
\large{Nadiya Zhutova,\textit{$^{a,b}$} Florent Réal,\textit{$^{b}$} Eric Renault,\textit{$^{c}$} Valérie Vallet,\textit{$^{b}$} and Rémi Maurice$^{\ast}$\textit{$^{a}$}}
\end{center}

\vspace{0.3cm}

\noindent\normalsize{Polonium (Po, $Z$ = 84) is a main-block element with poorly known physico-chemical properties. Not much information has been firmly acquired since its discovery by Marie and Pierre Curie in 1898, especially regarding its speciation in aqueous solution and spectroscopy. In this work, we revisit the absorption properties of two complexes, \ce{[PoCl5]-} and \ce{[PoCl6]^{2-}}, using quantum mechanical calculations. These complexes have the potential to exhibit a maximum absorption at \SI{418}{\nm} in HCl medium (for \SI{0.5}{\mol\per\liter} concentrations and above). Initially, we examine the electronic spectra of the \ce{Po^{4+}} free ion and of its isoelectronic analogue, \ce{Bi^{3+}}. In the spin-orbit configuration interaction (SOCI) framework. Our findings demonstrate that the SOCI matrix should be dressed with correlated electronic energies and that the quality of the spectra is largely improved by decontracting the reference states at the complete active space plus singles (CAS+S) level. Subsequently, we investigate the absorption properties of the \ce{[PoCl5]-} and \ce{[PoCl6]^{2-}} complexes in two stages. Firstly, we perform methodological tests at the MP2/def2-TZVP gas phase geometries, indicating that the decontraction of the reference states can there be skipped without compromising the accuracy significantly. Secondly, we study the solution absorption properties by means of single-point calculations performed at the solvated geometries, obtained by an implicit solvation treatment or a combination of implicit and explicit solvation. Our results highlight the importance of saturating the first coordination sphere of the \ce{Po^{IV}} ion to obtain a qualitatively correct picture. Finally, we conclude that the known-for-decades \SI{418}{\nm} peak could be attributed to a mixture of both the \ce{[PoCl5(H2O)]-} and \ce{[PoCl6]^{2-}} complexes. This finding not only aligns with the behaviour of the analogous \ce{Bi^{III}} ion under similar conditions but also potentially provides an explanation for previous discrepancies in the literature.} 

 \end{@twocolumnfalse} \vspace{0.3cm}

\vspace{0.6cm}

]

\footnotetext{\textit{$^{a}$~Univ Rennes, CNRS, ISCR (Institut des Sciences Chimiques de Rennes) -- UMR 6226, F-35000 Rennes, France; E-mail: remi.maurice@univ-rennes.fr}}
\footnotetext{\textit{$^{b}$~Univ. Lille, CNRS, UMR 8523-PhLAM-Physique des Lasers, Atomes et Molécules, F-59000 Lille, France; E-mail: valerie.vallet@univ-lille.fr}}
\footnotetext{\textit{$^{c}$~Nantes Université, CNRS, CEISAM UMR 6230, F-44000 Nantes, France}}
\footnotetext{\dag~Electronic Supplementary Information (ESI) available: Molecular structures (XYZ format) and examples of numerical compositions of the excited energy levels. }

\section{Introduction}

Polonium (Po, $Z$ = 84) is an unstable element displaying 33 radioactive isotopes with atomic masses ranging from 188 to 220 u. Discovered by Marie and Pierre Curie back in 1898, this main-group element remains quite secret. Its natural abundance is among the weakest ones on Earth, its concentration being for instance only about \SI{0.1}{ppb} in uranium ores\cite{bagnall1962chemistry}. Polonium isolation is complex, though possible, from ores. Nowadays, scientific studies are based on artificial polonium, which is obtained by irradiating bismuth targets with a neutron or proton beam. 

Alternatively, quantum mechanical calculations may be helpful for looking for trends in the periodic table and also for paving the way for future targeted experiments. In the last two decades, a couple of studies appeared in the literature. \citeauthor{Ayala2008po} focused on the hydrated and hydrolysed complexes of the solvated \ce{Po^{4+}} ion\cite{Ayala2008po, ayala2009general, ayala2010ab}, \citeauthor{rota2011zero} to the zero-field splitting of the \ce{Po2} hypothetical molecule\cite{rota2011zero}, \citeauthor{borschevsky2015ionization} on the electron affinity and on the ionization potential of the free Po atom\cite{borschevsky2015ionization}, and more recent studies were dedicated to chalcogen or hydrogen bonding with polonium \cite{zierkiewicz2019chalcogen, tulsiyan2022hydrogen}.

Besides those ground state studies, less effort has been put into the computation of electronic excited states and more specifically on the potential absorption properties of \ce{Po^{IV}} complexes. One may quote the work of \citeauthor{stoianov2019uv} on chlorinated complexes\cite{stoianov2019uv} and the work of \citeauthor{krishna2022metal} on a complex formed with the protocatechuic acid anion\cite{krishna2022metal}. From the first work, we have learnt that the excited states of \ce{Po^{IV}} complexes essentially expand on the 6p levels of Po, meaning that both scalar relativistic effects and the spin-orbit coupling must be properly accounted for. However, a few assumptions were made in this first study on the absorption properties of chlorinated \ce{Po^{IV}} complexes, which we aim to question in the current study. This study builds up also on our recent work on molecular geometries\cite{zhutova2022geometries}.

We will start by assessing the role of the decontraction of the perturbers in the $N$-electron valence state perturbation theory at second order (NEVPT2)\cite{Angeli:2001a,Angeli:2001b,Angeli:2002} and of the decontraction of the complete active space self-consistent field (CASSCF)\cite{Roos:1980} references for the subsequent spin-orbit configuration interaction (SOCI) treatment\cite{vallet2000two}. This study will be performed for both the free \ce{Po^{4+}} ion and for two representative binary complexes in the gas phase, namely \ce{[PoCl5]-} and \ce{[PoCl6]^{2-}}. After having selected an appropriate level for computing the excited energy levels, we will tackle the solvation issue, by an implicit solvation model, as in the previous study\cite{stoianov2019uv}, and by a more advanced combination of implicit and explicit solvation. This will allow us to reinterpret the respective absorption properties of the solvated \ce{[PoCl5]-} and \ce{[PoCl6]^{2-}} complexes and also to formulate conclusions on how to properly compute electronic absorption spectra of \ce{Po^{IV}} complexes.

\section{Methodology}

The correct computation of the electronic excited states at solvated geometries and in the presence of the SOC in \ce{Po^{IV}} complexes requires solving multiple issues related to electron correlation, relativity and solvation. Because of the single-reference characters of the ground states of such complexes, which has already been documented\cite{stoianov2019uv}, we have split the methodology section in two main parts, (i) geometry optimizations and (ii) electronic excited states. In each part, specific issues will be detailed. In particular, geometry optimizations may be performed in the gas phase or in the presence of an implicit or a combination of an implicit and explicit solvation model, and several levels of theory may be used to compute electronic excited states. Note that the \textit{Results and discussion} section will be conducted so as to sequentially address the electron correlation, relativity and solvation issues in systems of increasing complexity.

\subsection{Ground-state geometry optimizations}

All the geometry optimizations have been performed with the Gaussian 16 program package\cite{g16}.

\subsubsection{Gas phase structures}

Because these complexes correlate with the ground state of the closed-shell \ce{Po^{4+}} ion, the ground states of \ce{Po^{IV}} systems are generally also a closed-shell ones. Therefore, any single-reference method such as Møller–Plesset perturbation theory at second order (MP2)\cite{frisch1990direct,frisch1990semi,head1988mp2,saebo1989avoiding,head1994analytic}, coupled cluster with single, double and perturbative triple excitations (CCSD(T))\cite{purvis1982full,pople1987quadratic} or restricted Kohn-Sham density functional theory (DFT) with an appropriate exchange-correlation functional such as the almighty B3LYP \cite{Vosko:1980, lee1988development,Becke:1993, Stephens:1994} one should in principle be capable of converging on the right ground state and on qualitatively correct molecular structures by optimisation. 

In a previous work\cite{zhutova2022geometries}, we have revisited the computation of molecular geometries of hydrated and chlorinated complexes, complementing two previous studies from the literature\cite{Ayala2008po,stoianov2019uv}. We concluded that MP2 tends to slightly underestimate the bond distances involving Po whereas B3LYP tends to slightly overestimate them. Because of a small differential advantage in favour of MP2 (which underestimates less than B3LYP overestimates), we retained the MP2/def2-TZVP level of theory (with a combination of the ECP60MDF pseudopotential\cite{peterson2003b} and the associated def2-TZVP-PP basis set for Po and of the all-electron def2-TZVP\cite{weigend2005a} basis set for the lighter atoms). Following this work, we have applied the MP2/def2-TZVP level of theory for determining all the molecular geometries that are used in this work. Furthermore, we retained the same freezing scheme in the correlated calculations as in this previous work\cite{zhutova2022geometries} (\textit{i.e.} the 5s and 5p orbitals of Po are frozen). Note that the gas-phase structures of the bare \ce{[PoCl5]-} and \ce{[PoCl6]^{2-}} complexes are thus strictly identical to the previously reported ones. Also, the potential consequences of this choice will be commented in the \textit{Results and discussion} section.

\subsubsection{Solvated structures}

Two types of solvated structures have been determined, \textit{i.e.} structures of the bare \ce{[PoCl5]-} and \ce{[PoCl6]^{2-}} complexes under the influence of a conductor-like polarisable continuum model (C-PCM)\cite{Cossi:2003}, and structures for the microhydrated \ce{[PoCl5(H2O)_x]-} (x = 1,2) and \ce{[PoCl6(H2O)]^{2-}} complexes, in the gas phase and with the active C-PCM. Note that as a rule of thumb, it is necessary to stepwise increment the number of coordinated solvated molecules to determine the number of coordinated atoms in the first coordination sphere of the ion of interest (see for instance the recent work of Oher \textit{et al.}\cite{Oher:2023} and references therein), which explains why we considered the \ce{[PoCl5(H2O)2]-} and \ce{[PoCl6(H2O)]^{2-}} complexes in our study, even if we end up developing our reasoning more specifically on the \ce{[PoCl5]-}, \ce{[PoCl5(H2O)]-} and \ce{[PoCl6(H2O)]^{2-}} complexes, as will be later explained.

The C-PCM structures were determined at the same MP2/def2-TZVP level of theory. Because of the general lack of cavity parameters for the Po atom, we have used the default universal force field (UFF)\cite{Rappe:1992} value for the base radius of this atom, which is \SI{2.3545}{\angstrom}, and we chose to use those as well as for all the other atoms (\SIlist{1.9735;1.7500;1.4430}{\angstrom} for Cl, O and H, respectively). The $\alpha$ parameter has been left at its current default value in Gaussian, 1.1. The water solvent was considered ($\epsilon_r$ = 78.3553).

All the structures that are key to develop our scientific reasoning in the remainder of the article are given in the Supporting Information file.

\subsection{Electronic excited states}

The electronic excited states have been essentially computed by means of two codes, namely ORCA\cite{neese2018software} version~4.2.1 for the spin-orbit free part and EPCISO\cite{vallet2000two} for the part that involves the SOC. Note that the calculation of SOC integrals within the pseudopotential approximation is not implemented in ORCA, which is why even the c-SOCI calculations have been performed with EPCISO. Also, since EPCISO relies on MOLCAS\cite{MOLCAS}, the underlying spin-orbit free calculations have also been performed with this code.

\subsubsection{Spin-orbit free energies}

At the retained geometries, a series of single-point calculations are performed. Dependencies occur between those calculations. For a given system at a given geometry, the first calculation that is performed is a state-average (SA) CASSCF calculation. The same basis set as for the geometry optimisations is used. Thus, three main specific degrees of freedom emerge at this stage, (i) the size of the active space ($n$ electrons within $m$ orbitals), (ii) the number of roots that are computed and (iii) the SA scheme that is used. Following previous works\cite{real2006ab, stoianov2019uv}, we have used active spaces comprising 2 electrons in 4 orbitals. In the \ce{Bi^{3+}} and \ce{Po^{4+}} free ions, those correspond to the 6s and 6p valence orbitals, the ground state electronic structure being [\ldots]6s$^2$6p$^0$. In the studied chlorinated \ce{Po^{IV}} complexes, the active orbitals correspond to the fully bonding HOMO orbital of the ground state (which bears some 6s character) and to the essentially 6p orbitals of Po. From the ground state, three single excitations to the 6p$_x$, 6p$_y$ and 6p$_z$ orbitals are targeted, which result in spin-triplet or spin-singlet many-electron states\cite{stoianov2019uv}. Therefore, four spin-singlet and three spin-triplet roots are here computed by SA-CASSCF. We have used the default SA scheme in ORCA\cite{neese2012orca}, which consists in having equal weights for each block (for the $S$ = 0 and the $S$ = 1 roots, respectively), and equal weights for all the roots within a block. Therefore, each $S$ = 0 root weights \SI{12.5}{\percent} and each $S$ = 1 root \SI{16.7}{\percent} in the weighted sum of one-electron density matrices. 

The SA-CASSCF roots serve as references for the subsequent NEVPT2 calculations. Since the Dyall's zeroth-order Hamiltonian\cite{Dyall:1995} is used, the two-electron terms within the active electrons are considered, which generally prevents the occurrence of (spurious) intruder states. Two flavours are used in this work, namely the strongly-contracted (SC) NEVPT2 scheme\cite{Angeli:2001a,Angeli:2001b} and the partially contracted version of it\cite{Angeli:2002}, here via the fully internally contracted (FIC) implementation\cite{Sivalingam:2016} of ORCA. In both schemes, the SA-CASSCF states remain the references (internal contraction), so that both approaches only differ by the way the perturber functions are defined (external contraction), a larger function space being used in the FIC-NEVPT2 approach, leading in principle to more accurate results concomitant to an enhanced computational cost. Note that we do not report partially internally decontracted results that could be obtained with the quasi-degenerate (QD) formulation\cite{Angeli:2004} of SC-NEVPT2, simply because our set of spin-orbit free states do not allow for such partial decontraction in our reference systems (\textit{i.e.} the free ions and the bare \ce{[PoCl5]-} and \ce{[PoCl6]^{2-}} complexes, neither in the gas phase nor with the active C-PCM). Indeed, QD-NEVPT2 should only remix the SA-CASSCF roots if some of them belong to the same irreducible representation. Note that for computing the energy levels with EPCISO, the SA-CASSCF, SC-NEVPT2 and FIC-NEVPT2 electronic energies have been stored.

\subsubsection{Spin-orbit energy levels}

All the computations of the energy levels, \textit{i.e.} all the SOC calculations, have been performed with EPCISO\cite{vallet2000two}. This code starts over MOLCAS\cite{MOLCAS} which is used to compute both the integrals and also the SA-CASSCF orbitals. Unlike ORCA, MOLCAS uses a different set of SA orbitals for each block. In this work, we have used the SA-CASSCF orbitals of the $S$ = 1 block. EPCISO performs a configuration interaction (CI) (it can also add some perturbation theory), which can be done at both the spin-orbit free or at the SOCI level. 

In all the reported SOC calculations, the diagonal of the CI matrix was dressed with the previous ORCA energies, be them the SA-CASSCF, SC-NEVPT2 or the FIC-NEVPT2 ones. The SOC integrals were computed with the spin-dependent part of the ECP60MDF pseudopotential\cite{peterson2003b}. Two types of calculations were performed:

\begin{enumerate}
\item When the CASCI wave functions generated by EPCISO are readily used as references for the SOCI, the calculation is said as `contracted' (c-SOCI). In fact, this simply consists in diagonalising the $\bf{H}_{tot}= \bf{H}_{el}+\bf{H}_{SOC}$ matrix within the basis of the $M_S$ components of the retained spin-orbit free states. The resulting energy levels are thus expressed in this basis, and, for the sake of simplicity, we only report the compositions in terms of weights associated with the spin-orbit free states. In other words, the weights for the three components of each given spin-triplet spin-orbit free state are summed.
\item Alternatively, one may decontract the CASCI references within the CAS space (internal decontraction in this space according to the previously employed vocabulary) prior to performing the final SOCI step. In this case, the calculation is said `uncontracted' (uc-SOCI). The previous equation ($\bf{H}_{tot}= \bf{H}_{el}+\bf{H}_{SOC}$) still holds, but now the SOC matrix elements (which constitute $\bf{H}_{SOC}$) are computed in the basis of the $M_S$ components of the revised spin-orbit free states. Naturally, the way the decontraction is performed matters. In this work, we use the standard approach consisting in decontracting the CASCI states at the CAS plus single excitations (CAS+S) level, with no restrictions on the CAS+S space. If the CAS+S step were spin-orbit free, only orbital relaxation effects would be susceptible to be added (which is not likely to be the case here given the nature of the computed states). If this step includes both the spin-orbit free Hamiltonian and the SOC, both orbital relaxation effects and the spin-orbit polarisation are introduced (meaning that the interplay between electron correlation and the SOC is also there), which is the approach that has been used in the present work. By doing so, the SOC constant of a free 6p atom may be enhanced by up to $\sim$\SI{15}{\percent} (as it is the case for At\cite{Maurice:2015a}), leading to a much better description of the SOC and consequently to more accurate spectra.

\end{enumerate}

\section{Results and discussion}

\subsection{The isoelectronic \ce{Bi^{3+}} and \ce{Po^{4+}} free ions}

We start by discussing the results concerning the \ce{Bi^{3+}} and \ce{Po^{4+}} free ions. Studying the reference \ce{Po^{4+}} free ion will later be interesting in view of understanding the electronic structures of the \ce{Po^{IV}} complexes that are at the core of this work, especially in the \ce{[PoCl6]^{2-}} case. From a methodological point of view, the SOC is larger in the free ions than it is in related complexes. Therefore, a method that captures well the SOC in the free ions is also expected to be readily applicable to the complexes. For validation purposes, we have looked for an analogous for which reference experimental and computational data is available. This analogous must fulfill two criteria, (i) belong to the 6p family (the SOC being much smaller in the 5p block, not much is added from c-SOCI to uc-SOCI in this case) and (ii) be valence isoelectronic with the \ce{Po^{4+}} ion. It appears that the \ce{Bi^{3+}} ion fulfills these conditions and that Bi is an immediate neighbor of Po in the periodic table. In both cases (\ce{Bi^{3+}} and \ce{Po^{4+}}), we aim at describing the electronic excited states that result from single excitations from the ground state [\ldots]6s$^2$6p$^0$ electronic configuration to the single-excited [\ldots]6s$^1$6p$^1$ one. At the spin-orbit free level, those single-excited states correspond to the $^3$P and $^1$P terms, while the ground state is $^1$S. 

In view of the subsequent SOCI calculations, it is important to start with accurate spin-orbit free energies. As can be seen in \autoref{tab:1}, the excitation energies to the $^3$P and $^1$P excited terms are quite affected by the dynamic correlation that is introduced by NEVPT2. While $^3$P is pushed up by $\sim$\SI{1}{\electronvolt}, $^1$P is pushed down by about the same quantity, resulting in an average decrease of $\Delta$ = $E(^1\text{P})-E(^3\text{P})$ of $\sim$\SI{2}{\electronvolt} (in fact, 1.89 for \ce{Bi^{3+}} and 2.06 for \ce{Po^{4+}}). Therefore, it is already clear that the SA-CASSCF spin-orbit free energies cannot lead to accurate spectra for the targeted states. Therefore, we will only report below the SOCI results obtained with the SC-NEVPT2 and FIC-NEVPT2 spin-orbit free energies. Note that although SC-NEVPT2 and FIC-NEVPT2 lead to practically the same results, we retain both sets of spin-orbit free energies for dressing the SOCI matrices since such types of calculations have never been compared in this context.

\begin{table}[H]
\small
  \caption{\ Spin-orbit free excitation energies (in \si{\electronvolt}) from the $^1$S ground state of the  of the \ce{Bi^{3+}} and \ce{Po^{4+}} free ions. The SA-CASSCF references are built in each case on a 2 electrons in 4 orbitals active space. }
  \label{tab:1}
  \begin{tabular*}{0.48\textwidth}{@{\extracolsep{\fill}}lllll}
    \toprule
    System & Term & SA-CASSCF & SC-NEVPT2 & FIC-NEVPT2 \\
    \midrule
    \multirow{2}{*}{\ce{Bi^{3+}}} & $^3$P & 10.14 & 11.15 & 11.14 \\
    & $^1$P & 14.57 & 13.69 & 13.64 \\
    \midrule    
    \multirow{2}{*}{\ce{Po^{4+}}} & $^3$P & 12.10 & 13.09 & 13.09 \\
    & $^1$P & 16.99 & 15.92 & 15.87 \\
    \bottomrule
  \end{tabular*}
\end{table}

When the SOC is introduced, $^3$P splits in three energy levels, here labelled according to a Russell-Saunders coupling scheme ($J=L+S$). If only the first-order SOC of this $^3$P term were considered (Russell-Saunders coupling scheme), the four theoretical excitation energies of interest would be:
\begin{align}
\label{eq:1}
    E(^3\text{P}_0) & = E(^3\text{P}) - 2\lambda\\
\label{eq:2}
    E(^3\text{P}_1)&= E(^3\text{P}) - \lambda\\
\label{eq:3}
    E(^3\text{P}_2) &= E(^3\text{P}) + \lambda\\
\label{eq:4}
    E(^1\text{P}_1) &= E(^1\text{P}) 
\end{align}
\noindent where $\lambda$ is the SOC constant (positive defined). As can be seen in \autoref{tab:2}, this Russell-Saunders picture is perfectible. In fact, $E(^3\text{P}_1)$ and $E(^1\text{P}_1)$ are coupled at the second order of perturbation by the SOC operator. A more refined theoretical picture thus requires at least an \textit{ad hoc} correction to eqns. \ref{eq:2} and \ref{eq:4}, \textit{e.g.}:
\begin{align}
\label{eq:5}
    E(^3\text{P}_1) &= E(^3\text{P}) - \lambda - \frac{x}{\Delta E}\\
\label{eq:6}
    E(^1\text{P}_1) &= E(^1\text{P}) + \frac{x}{\Delta E}
\end{align}
\noindent where $x$ is a positive defined and remains here an \textit{ad hoc} parameter (it will not be further developed), and $\Delta E$, the energy spacing between the $^3$P and $^1$P spin-orbit free states. When eqs. \ref{eq:1}, \ref{eq:3}, \ref{eq:5} and \ref{eq:6} are considered, all the c-SOCI results reported in \autoref{tab:2} are now well explained. Indeed, one can extract the $\lambda$ SOC constants by making use of eqns. \ref{eq:1} and \ref{eq:3} (see also \autoref{tab:3}), and also the $x$ \textit{ad hoc} parameters, and then, reconstruct the excitation energies with high accuracy (not shown), which is not surprising owing to the chosen configuration space: only the theoretically described first- and second-order SOCs are here active in the calculations. 

\begin{table}[H]
\small
  \caption{\ Excitation energies (in eV) from the  $^1$S$_0$ ground state of the energy levels of the \ce{Bi^{3+}} and \ce{Po^{4+}} free ions. The SA-CASSCF references are built in each case on a 2 electrons in 4 orbitals active space. The mean absolute error (MAE) is computed with respect to the experimental reference set.}
  \label{tab:2}
  \begin{tabular*}{0.48\textwidth}{@{\extracolsep{\fill}}lllllll}
    \toprule
    \multirow{2}{*}{System} & \multirow{2}{*}{Level} & \multicolumn{2}{c}{SC-NEVPT2} & \multicolumn{2}{c}{FIC-NEVPT2} & \multirow{2}{*}{Expt.\cite{Expt1933}} \\
    \cmidrule(l){3-4}\cmidrule(l){5-6}
    &  & c-SOCI & uc-SOCI & c-SOCI & uc-SOCI & \\
    \midrule
    \multirow{4}{*}{\ce{Bi^{3+}}} & $^3$P$_0$ & 9.30 & 8.90 & 9.30 & 8.90 & 8.80 \\
    & $^3$P$_1$ & 9.79 & 9.51 & 9.78 & 9.50 & 9.41 \\
    & $^3$P$_2$ & 12.08 & 11.99 & 12.07 & 11.98 & 11.95 \\
    & $^1$P$_1$ & 14.13 & 13.96 & 14.09 & 13.92 & 14.21 \\
    \\
    \multicolumn{2}{l}{MAE} & 0.27 & 0.12 & 0.28 & 0.13 & \\
    \midrule    
    \multirow{4}{*}{\ce{Po^{4+}}} & $^3$P$_0$ & 10.56 & 10.09 & 10.56 & 10.09 & -- \\
    & $^3$P$_1$ & 11.16 & 10.81 & 11.15 & 10.80 & -- \\
    & $^3$P$_2$ & 14.37 & 14.27 & 14.37 & 14.26 & -- \\
    & $^1$P$_1$ & 16.60 & 16.39 & 16.56 & 16.35 & -- \\
    \bottomrule
  \end{tabular*}
\end{table}

\begin{table}[H]
\small
  \caption{\ Polyelectronic ($\lambda$) spin-orbit coupling constants (in eV) of the excited \ce{Bi^{3+}} and \ce{Po^{4+}} free ions obtained from the $^3$P$_2-^3$P$_0$ energy differences. The SA-CASSCF references are built in each case on a 2 electrons in 4 orbitals active space.}
  \label{tab:3}
  \begin{tabular*}{0.48\textwidth}{@{\extracolsep{\fill}}llllll}
    \toprule
    \multirow{2}{*}{System} & \multicolumn{2}{c}{SC-NEVPT2} & \multicolumn{2}{c}{FIC-NEVPT2} & \multirow{2}{*}{Expt.\cite{Expt1933}} \\
    \cmidrule(l){2-3}\cmidrule(l){4-5}
    & c-SOCI & uc-SOCI & c-SOCI & uc-SOCI & \\
    \midrule
    {\ce{Bi^{3+}}} & 0.93 & 1.03 & 0.92 & 1.03 & 1.05 \\
    {\ce{Po^{4+}}} & 1.27 & 1.39 & 1.27 & 1.39 & -- \\
    \bottomrule
  \end{tabular*}
\end{table}

In the \ce{Bi^{3+}} free ion, for which we benefit from accurate experimental data, it is interesting to compare the respective accuracies of the c-SOCI and uc-SOCI approaches. From \autoref{tab:2}, one can see that the mean-absolute error (MAE) is decreased by more than a factor of two when going from c-SOCI to uc-SOCI, independent from the chosen NEVPT2 flavour (SC or FIC). In both cases, a MAE of $\sim$\SI{0.1}{\electronvolt} is observed, meaning that satisfactory accuracy is reached. This tends to validate both the uc-SOCI method overall and also the dressing of the uc-SOCI matrix with either the SC-NEVPT2 or FIC-NEVPT2 spin-orbit free energies. Note that the error that we have observed at the c-SOCI levels is larger but comparable to the one reported by Roos and Malmqvist\cite{Roos:2004b} for the Po free atom ($\sim$0.2 eV), for which the SOC is meant to be smaller than in our case (smaller effective charge), which again illustrates the importance of the spin-orbit polarisation for 6p atoms and ions.

To further assess the quality of our uc-SOCI results, we have then compared the computed $\lambda$ constant for the \ce{Bi^{3+}} ion to the experimental one, also derived by using eqs. \ref{eq:1} and \ref{eq:3}. This constant is enhanced by \SI{12}{\percent} by using the uc-SOCI approach, to reach a very close agreement with the experimental one. One can thus conclude that our representation of the SOC is especially accurate at the uc-SOCI level and that then the remaining errors that are observed in \autoref{tab:2} may essentially be attributed to some lack of electron correlation. 

At a finer level of description, it is worth noting that eqs. \ref{eq:1}, \ref{eq:3}, \ref{eq:5} and \ref{eq:6} cannot fully reproduce the uc-SOCI energies after extraction of the $\lambda$ and $x$ parameters. We believe that this is due to the revision of the references at the CAS+S level, which may introduce more subtle interactions with other excited configurations\cite{Martin:1972}.

In this work, we have relied on the ECP60MDF pseudopotential. It is worth commenting on this choice here for the sake of completeness. First, it was shown by R\'eal \textit{et al.}\cite{real2006ab} that it is important to use a good quality pseudopotential such as the ECP60MDF one to improve accuracy (as compared with the larger-core ECP78MWB\cite{Kuchle:1991} and ECP78MDF\cite{Stoll:2002} ones). Second, test calculations have shown that similar quality results can be obtained with a good all-electron basis set such as the ANO-RCC-VTZP one\cite{Roos:2004}. Third and last, it is clear from Tables \autoref{tab:2} and \autoref{tab:3} that our pseudopotential based results are in good agreement with experiment.

After having validated once more\cite{real2006ab} the uc-SOCI method to compute the spectra of the \ce{Bi^{3+}} ion, it is interesting to compare the \ce{Bi^{3+}} and \ce{Po^{4+}} results. First, the \ce{Po^{4+}} free ion qualitatively behaves in a similar way as the \ce{Bi^{3+}} one, meaning that the previous intermediate conclusion still holds (the NEVPT2 energies are more accurate than the SA-CASSCF ones, the uc-SOCI method significantly improve over the c-SOCI one, both the SC-NEVPT2 and FIC-NEVPT2 electronic energies are similar, \textit{etc.}). Quantitatively, a few points are worth noting:

\begin{enumerate}
\item The $\lambda$ constant is relatively less enhanced by the uc-SOCI method (+\SI{9}{\percent}) than in \ce{Bi^{3+}}, even if the absolute enhancement in terms of eV is slightly more important.
\item The SOC constant of the \ce{Po^{4+}} free ion is significantly larger than the one of the isoelectronic \ce{Bi^{3+}} ion. We simply attribute this to the larger effective charge.
\item The spin-orbit free energy difference between the $^3$P and $^1$P terms ($\Delta E$) is larger in the \ce{Po^{4+}} case (2.8 \textit{vs.} \SI{2.5}{\electronvolt} at the NEVPT2 levels) than in the \ce{Bi^{3+}} case.
\end{enumerate}

Apart from the above-mentioned points, we observe very similar behaviours, which further justifies the retained analogy with the \ce{Bi^{3+}} free ion for methodological purposes.

\subsection{The bare \ce{[PoCl5]-} and \ce{[PoCl6]^{2-}} complexes in the gas phase}

Before tackling the solvation, it is interesting to focus on the \ce{[PoCl5]-} and \ce{[PoCl6]^{2-}} complexes in the gas phase. This intermediate step will allow us to derive methodological conclusions that are free of solvation model issues and in particular see if the conclusions that were drawn from the free-ion study are readily transferable to the \ce{Po^{IV}} complexes or if additional subtleties or simplifications apply. As already reported elsewhere\cite{stoianov2019uv,zhutova2022geometries}, the structures of the bare \ce{[PoCl5]-} and \ce{[PoCl6]^{2-}} complexes in the gas phase belong to $D_{3h}$ and $O_{h}$ symmetry point groups (SPGs), respectively. As a consequence, excitations to the 6p$_{x,y,z}$ levels generate two many-electron spin-orbit free states per spin multiplicity in the \ce{[PoCl5]-} complex (of $E^\prime$ and $A_2^{\prime\prime}$ symmetries), and only one in the \ce{[PoCl6]^{2-}} one ($T_{1u}$). The SA-CASSCF, SC-NEVPT2 and FIC-NEVPT2 spin-orbit free excitation energies are reported in \autoref{tab:4} and one can already observe a strong down shift (about 6~eV) of the electronic transitions in the complexes compared to the isolated ion.

\begin{table}[H]
\small
  \caption{\ Spin-orbit free excitation energies (in eV) from the ground singlet states of the bare \ce{[PoCl5]-} and \ce{[PoCl6]^{2-}} complexes in the gas phase. The SA-CASSCF references are built in each case on a 2 electrons in 4 orbitals active space. }
  \label{tab:4}
  \begin{tabular*}{0.48\textwidth}{@{\extracolsep{\fill}}llllll}
    \toprule
    System & SPG & State & SA-CASSCF & SC-NEVPT2 & FIC-NEVPT2 \\
    \midrule
    \multirow{4}{*}{\ce{[PoCl5]-}} & \multirow{4}{*}{$D_{3h}$} & $^3E^\prime$ & 3.58 & 3.56 & 3.53\\
    & & $^3A_2^{\prime\prime}$ & 4.58 & 4.47 & 4.45 \\
    & & $^1E^\prime$ & 5.04 & 3.95 & 3.89\\
    & & $^1A_2^{\prime\prime}$ & 6.38 & 5.03 & 4.95 \\
    \midrule    
    \multirow{2}{*}{\ce{[PoCl6]^{2-}}} & \multirow{2}{*}{$O_{h}$} & $^3T_{1u}$ & 4.50 & 4.34 & 4.32 \\
    & & $^1T_{1u}$ & 5.97 & 4.76 & 4.71 \\
    \bottomrule
  \end{tabular*}
\end{table}

In the case of the \ce{[PoCl5]-} complex, two main effects are observed,  the correlation that is introduced at the NEVPT2 levels (i) reduces the difference between the mean energy of the excited spin-triplet states and spin-singlet states by more than a factor of 3 and (ii) the order of the spin-free states changes ($^1E^\prime$ becomes lower in energy than $^3A_2^{\prime\prime}$). A similar situation occurs in the \ce{[PoCl6]^{2-}} complex, apart from the state ordering inversion, the single-triplet energy difference being reduced by a factor of 3.5 at the SC-NEVPT2 level and 3.8 at the FIC-NEVPT2 one. As for the \ce{Po^{4+}} free ion, it is thus crucial to use correlated energies for the forthcoming SOCI calculations.

In \autoref{tab:5}, SOCI excited energies are reported. In the \ce{[PoCl5]-} case, the 12 excited roots expand on 8 energy levels, some of them being doubly degenerate ($E^\prime$ and $E^{\prime\prime}$ symmetries). The generic compositions in terms of summed weights over the spin components of each spin-orbit free states are given in \autoref{tab:5}. The numerical weights  that are obtained at the c-SOCI level together with the SC-NEVPT2 are given in Table S1. Note that for the sake of predicting UV-Vis spectra, only the transitions leading to energy levels with non-zero components on the spin-singlet states are active. For instance, a transition to the first excited energy level ($A_1^{\prime\prime}$ ) is not allowed, while transitions to the second ($A_2^{\prime\prime}$) and third ($E^{\prime}$) ones are. If we specifically focus on these two excited energy levels, it is interesting to note that the c-SOCI result with the SC-NEVPT2 energies is very close to the uc-SOCI one with the FIC-NEVPT2 energies. Moreover, the SC-NEPVT2 and FIC-NEVPT2 results are also pretty close one another, as are the c-SOCI and uc-SOCI ones. As a consequence, computation of the lowest-lying energy levels can be safely done at the c-SOCI level together with the SC-NEVPT2 energies.

\begin{table*}[h]
\small
  \caption{\ Excitation energies (in eV) from the ground $^1A_1^{\prime}$ and $^1A_{1g}$ energy levels of the bare \ce{[PoCl5]-} and \ce{[PoCl6]^{2-}} complexes in the gas phase. The SA-CASSCF references are built in each case on a 2 electrons in 4 orbitals active space. The compositions are given in terms of the states of \autoref{tab:4}.}
  \label{tab:5}
  \begin{tabular*}{\textwidth}{@{\extracolsep{\fill}}llllllll}
    \toprule
    \multirow{2}{*}{System} & \multirow{2}{*}{SPG} & \multirow{2}{*}{Symmetry} & \multirow{2}{*}{Composition} & \multicolumn{2}{c}{SC-NEVPT2} & \multicolumn{2}{c}{FIC-NEVPT2} \\
    \cmidrule(l){5-6}\cmidrule(l){7-8}
    & & & & c-SOCI & uc-SOCI & c-SOCI & uc-SOCI \\
    \midrule
    \multirow{8}{*}{\ce{[PoCl5]-}} & \multirow{8}{*}{$D_{3h}$} & $A_1^{\prime\prime}$ & $a*^3E^\prime+(1-a)*^3A_2^{\prime\prime}$ & 2.46 & 2.48 & 2.44 & 2.46 \\
    & & $A_2^{\prime\prime}$ & $b*^3E^\prime+(1-b)*^1A_2^{\prime\prime}$ & 2.55 & 2.57 & 2.52 & 2.54 \\
    & & $E^{\prime}$ & $\omega_1*^3E^\prime+\omega_2*^3A_2^{\prime\prime}+\omega_3*^1E^\prime$ & 2.60 & 2.62 & 2.57 & 2.58 \\
    & & $E^{\prime\prime}$ & $^3E^\prime$ & 4.20 & 4.18 & 4.18 & 4.16 \\
    & & $E^{\prime}$ & $\omega_1^\prime*^3E^\prime+\omega_2^\prime*^3A_2^{\prime\prime}+\omega_3^\prime*^1E^\prime$ & 4.40 & 4.39 & 4.36 & 4.35 \\
    & & $A_1^{\prime\prime}$ & $(1-a)*^3E^\prime+a*^3A_2^{\prime\prime}$ & 4.94 & 4.93 & 4.92 & 4.91 \\
    & & $E^{\prime}$ & $\omega_1^{\prime\prime}*^3E^\prime+\omega_2^{\prime\prime}*^3A_2^{\prime\prime}+\omega_3^{\prime\prime}*^1E^\prime$ & 5.00 & 4.99 & 4.96 & 4.96 \\
    & & $A_2^{\prime\prime}$ & $(1-b)*^3E^\prime+b*^1A_2^{\prime\prime}$ & 5.41 & 5.41 & 5.34 & 5.33 \\
    \midrule    
    \multirow{4}{*}{\ce{[PoCl6]^{2-}}} & \multirow{4}{*}{$O_{h}$} & $A_{1u}$ & $^3T_{1u}$ & 2.75 & 2.77 & 2.73 & 2.76 \\
    & & $T_{1u}$ & $c*^3T_{1u}+(1-c)*^1T_{1u}$ & 2.88 & 2.90 & 2.85 & 2.88 \\
    & & $T_{2u}\oplus E_{u}$ & $^3T_{1u}$ & 5.14 & 5.13 & 5.12 & 5.11 \\
    & & $T_{1u}$ & $(1-c)*^3T_{1u}+c*^1T_{1u}$ & 5.44 & 5.43 & 5.39 & 5.38 \\
    \bottomrule
  \end{tabular*}
\end{table*}

In the \ce{[PoCl6]^{2-}} case, the situation is much simpler: two energy levels correlate only with $^3T_{1u}$, meaning that the energy difference between those two energy levels can be readily used to derive the SOC constant in this complex:

\begin{equation}
    \label{eq:7}
\lambda = \frac{E(T_{2u}\oplus E_{u})-E(A_{1u})}{3}
\end{equation}

\noindent where $\lambda$ is expected to be reduced by covalency compared to the free ion one. Also, the two remaining excited energy levels, of $T_{1u}$ symmetry, are characterized by a mixture of spin-triplet and spin-singlet characters (those correlate with $^3P_1$ and $^1P_1$ in the reference free ion). Here, the c-SOCI result obtained with the SC-NEVPT2 energies perfectly matches the uc-SOCI one with the FIC-NEVPT2 energies for the second excited energy level ($T_{1u}$). As explained in a previous publication \cite{stoianov2019uv}, the essentially spin-triplet energy levels that are $\sim$\SI{3}{\electronvolt} above the ground state are the ones that we are targeting, since these translate in a maximum of absorption near the experimental value of 418 nm \cite{moyer1956polonium}. We thus conclude that the c-SOCI/SC-NEVPT2 level is very accurate for our purposes (\textit{vide infra}), because of (small) error compensations (both the studied decontractions have opposite consequences on the excitation energies of interest). 

As a final validation of the c-SOCI/SC-NEVPT2 level for the rest of the article, we have reported the SOC constants that are obtained at the four main levels of interest in \autoref{tab:6}. In fact, all the SOC constants are in good agreement, meaning that our representation of the SOC is already quantitative at the c-SOCI/SC-NEVPT2 level in this complex and thus supposedly in other similar complexes such as \ce{[PoCl5]-} (\autoref{tab:5} being in line with this), contrary to the \ce{Po^{4+}} case for which the spin-orbit polarisation was much more important. In fact, the SOC is quite significantly reduced in the \ce{[PoCl6]^{2-}} case, with a reduction factor of $\sim$0.6. This is indicative of a significant covalent character in the Po--Cl bonds of interest for these two excited energy levels. Though we have not performed actual bonding analyses for these levels, we note that this is somehow consistent with the significant delocalisation indices that we have previously reported for the ground state (0.5, which would be indicative of ``half-bonds'')\cite{zhutova2022geometries}. 

\begin{table}[H]
\small
  \caption{\ Polyelectronic ($\lambda$) spin-orbit coupling constant (in eV) of the excited bare \ce{[PoCl6]^{2-}} complex obtained from the $(T_{2u}\oplus E_{u})-A_{1u}$ energy difference. The SA-CASSCF references are built in each case on a 2 electrons in 4 orbitals active space.}
  \label{tab:6}
  \begin{tabular*}{0.48\textwidth}{@{\extracolsep{\fill}}cccc}
    \toprule
    \multicolumn{2}{c}{SC-NEVPT2} & \multicolumn{2}{c}{FIC-NEVPT2} \\
    \cmidrule(l){1-2}\cmidrule(l){3-4}
    c-SOCI & uc-SOCI & c-SOCI & uc-SOCI \\
    \midrule
    0.80 & 0.79 & 0.80 & 0.78 \\
    \bottomrule
  \end{tabular*}
\end{table}

\begin{figure}[H]
\centering
  \includegraphics[width=8cm]{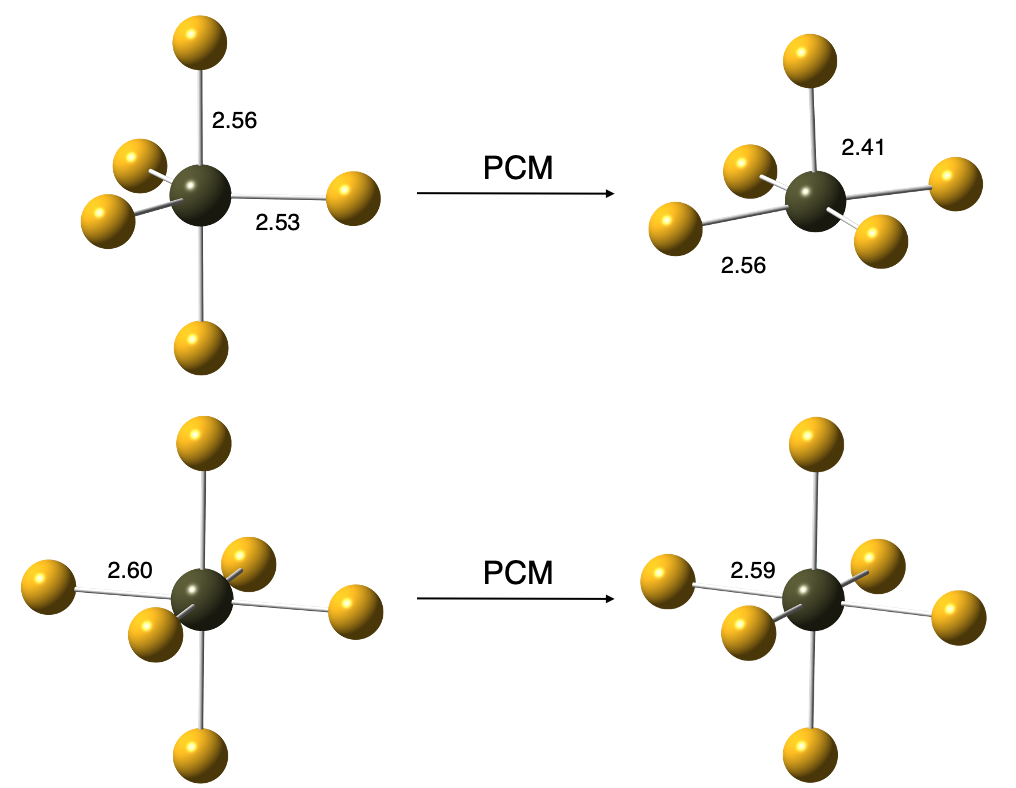}
  \caption{Representations of the molecular geometries of the bare \ce{[PoCl5]-} and \ce{[PoCl6]^{2-}} complexes in the gas phase (left) and under the influence of the polarisable continuum model (right). Colour code: polonium (dark green), chlorine (yellow).}
  \label{fgr:1}
\end{figure}

\subsection{The solvated \ce{[PoCl5]-} and \ce{[PoCl6]^{2-}} complexes}

\subsubsection{Implicit solvation}

We start our analysis with the structures that have been obtained under the influence of the C-PCM. As can be seen in \autoref{fgr:1}, the structure of the bare \ce{[PoCl5]-} complex drastically changes from $D_{3h}$ to $C_{4v}$, while the structure of the bare \ce{[PoCl6]^{2-}} complex is qualitatively maintained ($O_h$). In \ce{[PoCl5]-}, this symmetry change translates into an inversion of the $E$ and $A$ spin-orbit free states (see Tables \ref{tab:4} and \ref{tab:7}), simply because it becomes more favourable to excite to the 6p$_z$ orbital \textit{vs.} the degenerate 6p$_{x,y}$ ones. The crystal-field splitting is still of the order of \SI{1}{\electronvolt}, with significant deviations between the value derived from the spin-triplet states, $|E(^3 E)-E(^3 A)|$, and the one derived from the spin-singlet ones. A fine understanding of these differences is out of the scope of the present work, we only require accurate spin-orbit free energies to further process with the c-SOCI step. 

In the \ce{[PoCl6]^{2-}} complex, the symmetry is not broken by the C-PCM, meaning that the situation is quite similar. The singlet-triplet splitting is practically untouched (\SI{0.42}{\electronvolt} in the gas phase and \SI{0.43}{\electronvolt} with the C-CPM), while the mean single excitation energy is enhanced from \SIrange{4.55}{4.85}{\electronvolt} at the SC-NEVPT2 level, which is in line with the shorter Po--Cl bonds (the 6p orbitals display antibonding characters with respect to the ligands). 

\begin{table}[H]
\small
  \caption{\ Spin-orbit free excitation energies (in eV) from the ground singlet states of the \ce{[PoCl5]-} and \ce{[PoCl6]^{2-}} complexes at the solvated geometries (polarisable continuum model). The SA-CASSCF references are built in each case on a 2 electrons in 4 orbitals active space.}
  \label{tab:7}
  \begin{tabular*}{0.48\textwidth}{@{\extracolsep{\fill}}llllll}
    \toprule
    System & SPG & State & SC-NEVPT2 \\
    \midrule
    \multirow{4}{*}{\ce{[PoCl5]-}} & \multirow{4}{*}{$C_{4v}$} & $^3A_1$ & 3.39 \\
    & & $^3E$ & 4.77 \\
    & & $^1A_1$ & 3.61 \\
    & & $^1E$ & 5.43 \\
    \midrule    
    \multirow{2}{*}{\ce{[PoCl6]^{2-}}} & \multirow{2}{*}{$O_{h}$} & $^3T_{1u}$ & 4.64 \\
    & & $^1T_{1u}$ & 5.07 \\
    \bottomrule
  \end{tabular*}
\end{table}

The excitation energies obtained after the introduction of the SOC are displayed in \autoref{tab:8}. Following the change in the crystal field, the energy levels of the \ce{[PoCl5]-} complex do not show up in the same order as in the gas phase, with notable inversions of doubly-degenerate \textit{vs.} non-degenerate levels. Other features are maintained, such as the occurrence of energy levels of pure spin-triplet characters, and the presence of two very close energy levels $\sim$3 eV above the ground state with non-zero spin-singlet character and thus non-zero oscillator strengths. While the energy difference between those two levels is the same at the gas phase structure and at the C-PCM one (0.05 eV), the mean excitation energy is pushed up by nearly 0.4 eV. Since in the C-PCM structure there is room for at least one more coordination bond, we will study the combination of implicit and explicit solvation. 

\begin{table*}[h]
\small
  \caption{\ Excitation energies (in eV) from the $^1A_1$ and $^1A_{1g}$ ground energy levels of the \ce{[PoCl5]-} and \ce{[PoCl6]^{2-}} complexes at the solvated geometries (polarisable continuum model). The SA-CASSCF references are built in each case on a 2 electrons in 4 orbitals active space. The compositions are given in terms of the states of \autoref{tab:7}.}
  \label{tab:8}
  \begin{tabular*}{\textwidth}{@{\extracolsep{\fill}}lllll}
    \toprule
    \multirow{2}{*}{System} & \multirow{2}{*}{SPG} & \multirow{2}{*}{Symmetry} & \multirow{2}{*}{Composition} & \multicolumn{1}{c}{SC-NEVPT2} \\
    & & & & c-SOCI \\
    \midrule
    \multirow{8}{*}{\ce{[PoCl5]-}} & \multirow{8}{*}{$C_{4v}$} & $A_2$ & $a*^3A_1+(1-a)*^3E$ & 2.84 \\
    & & $E$ & $\omega_1*^3A_1+\omega_2*^3E+\omega_3*^1E$ & 2.93 \\
    & & $A_1$ & $b*^3E+(1-b)*^1A_1$ & 2.98 \\
    & & $A_2$ & $(1-a)*^3A_1+a*^3E$ & 4.64 \\
    & & $A_1$ & $(1-b)*^3E+b*^1A_1$ & 4.72 \\
    & & $E$ & $\omega_1^\prime*^3A_1+\omega_2^\prime*^3E+\omega_3^\prime*^1E$ & 4.80 \\
    & & $B_1\oplus B_2$ & $^3E$ & 5.51 \\
    & & $E$ & $\omega_1^{\prime\prime}*^3A_1+\omega_2^{\prime\prime}*^3E+\omega_3^{\prime\prime}*^1E$ & 5.92 \\
    \midrule    
    \multirow{4}{*}{\ce{[PoCl6]^{2-}}} & \multirow{4}{*}{$O_{h}$} & $A_{1u}$ & $^3T_{1u}$ & 3.06 \\
    & & $T_{1u}$ & $c*^3T_{1u}+(1-c)*^1T_{1u}$ & 3.19 \\
    & & $T_{2u}\oplus E_{u}$ & $^3T_{1u}$ & 5.45 \\
    & & $T_{1u}$ & $(1-c)*^3T_{1u}+c*^1T_{1u}$ & 5.75 \\
    \bottomrule
  \end{tabular*}
\end{table*}

In the case of the \ce{[PoCl6]^{2-}} complex, the use of the C-PCM structure unsurprisingly leads to a similar spectrum with all the computed excited energy levels being relatively destabilised with respect to the ground state, which is a direct consequence of what was observed at the spin-orbit free level. The energy level of interest for the final discussion is the energy level of dominant spin-triplet and significant spin-singlet character, \textit{i.e.} the first $T_{1u}$ level. The corresponding excitation energy is enlarged from 2.88 to 3.19 eV at the c-SOCI/SC-NEVPT2 level, which is concomitant with a bond shortening of only 0.015 {\AA}! We must thus stress that our excitation energies suffer from significant uncertainty here, because alternative choices for the underlying method used for the geometry optimization and/or for the radius used in the C-CPM step may result in quite different excitation energies. Since here we do not want to force any artificial agreement with experiment (\textit{vide infra}) and since we are more interested in comparing the absorption properties of the two closely-related \ce{[PoCl5]-} and \ce{[PoCl6]^{2-}} complexes, we fix both the reference level of theory to determine the geometries to the MP2/def2-TZVP one and the reference level of theory for computing the excitation energies to the c-SOCI/SC-NEVPT2 one and only pursue by looking at differential effects between both the systems.

\subsubsection{Implicit and explicit solvation}

\begin{figure}[H]
\centering
  \includegraphics[width=8cm]{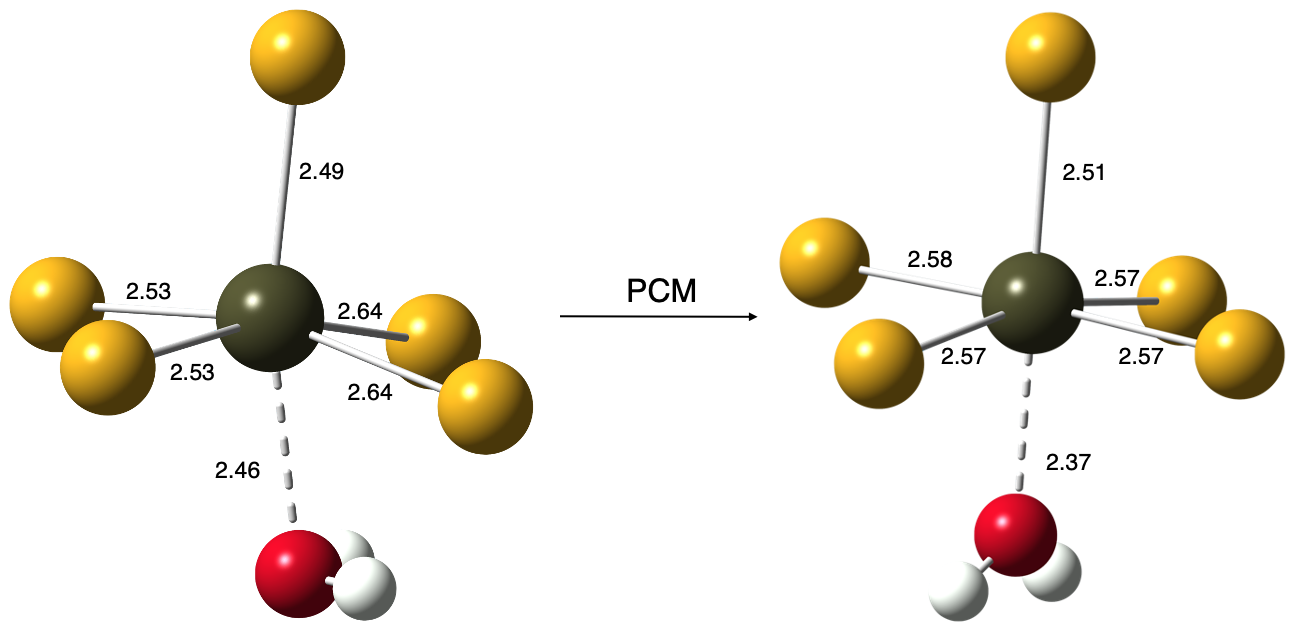}
  \caption{Representations of the molecular geometry of the hydrated \ce{[PoCl5(H2O)]-} complex in the gas phase and under the influence of the polarisable continuum model. Colour code: polonium (dark green), chlorine (yellow), oxygen (red) and hydrogen (white).}
  \label{fgr:2}
\end{figure}

In order to check the saturation of the first coordination sphere of the \ce{Po^{IV}} ion, we have considered the stepwise addition of water molecules in the quantum chemical treatment. The key structures in the gas phase and under the influence of the C-PCM are given in Supporting Information. In the case of the \ce{[PoCl5]-} complex, while the first water molecule enters the first coordination sphere independent of the phase (\textit{i.e.} with or without the C-PCM), the second one is clearly in the second sphere. Therefore, we retain the \ce{[PoCl5(H2O)]-} system for the forthcoming calculations. It is worth noting that the positions of the hydrogen atoms are qualitatively different in the absence or presence of the C-PCM: while they hydrogen-bind to chlorine atoms in the first case, they point outside the complex when the C-PCM is applied (see \autoref{fgr:2}). Consequently, with the aim of explaining data in solution, we only retain the C-PCM structure. This structure, though of formal $C_1$ symmetry, is close to be $C_{2v}$. Since in both cases no degenerate irreducible representation is available, we end up anyway with 12 distinct excitation energies (see for instance Table S3, corresponding to the straight $C_1$ optimized structure).

In the case of the \ce{[PoCl6]^{2-}} complex, even the first water molecule sits in the second coordination sphere of the \ce{Po^{IV}} ion. If only one water molecule is considered, the symmetry cannot be as high as $O_h$ anymore. This is why we prefer to keep the symmetry and only consider a complete first coordination sphere, as in the \ce{[PoCl5(H2O)]-} complex, and thus keep the \ce{[PoCl6]^{2-}} complex for the final discussion.

In the \ce{[PoCl5(H2O)]-} system, the second, third and fourth excited energy levels are the ones which correlate with the second and third one in the bare \ce{[PoCl5]-} complex. Because of the obtained geometry for the \ce{[PoCl5(H2O)]-} system, we may assess the influence of the added water molecular on the spectrum if we compare it to the one for the \ce{[PoCl5]-} complex under the influence of the C-PCM. The three excitation energies of interest in the \ce{[PoCl5(H2O)]-} system are 3.28, 3.36 and 3.37 eV (see Table S3), to be compared to the 2.93 and 2.98 ones for the bare \ce{[PoCl5]-} complex (see \autoref{tab:8}). Though the splitting between those excited energy levels is of the same order of magnitude, the mean excitation energy is pushed up by the coordinated water. Furthermore, this mean excitation energy of interest now is larger than the one observed in \ce{[PoCl6]^{2-}} (3.19 eV) and in fact closer to it. 

\begin{figure*}[h]
\centering
  \includegraphics[width=0.5\textwidth]{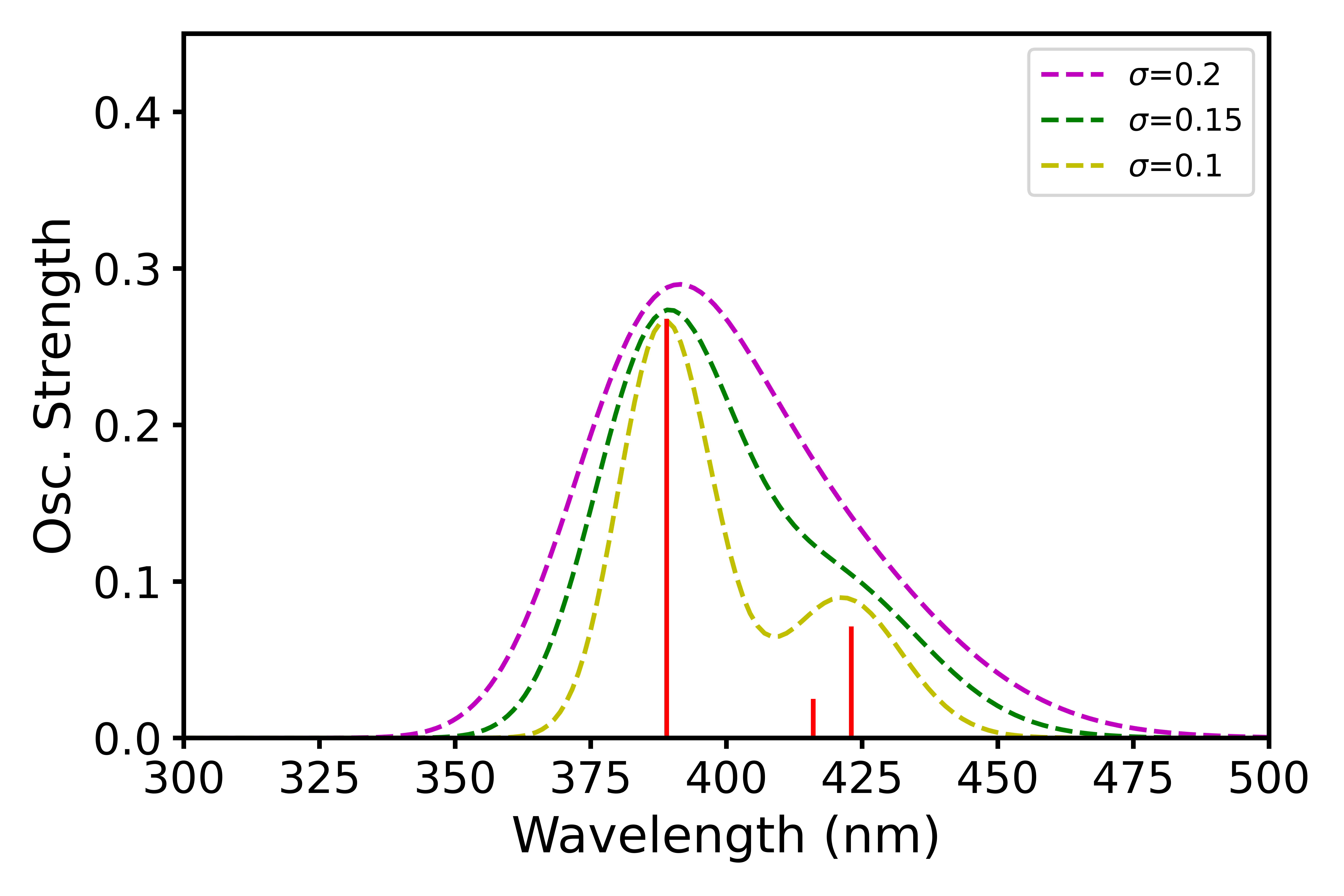}\hfill  \includegraphics[width=0.495\textwidth]{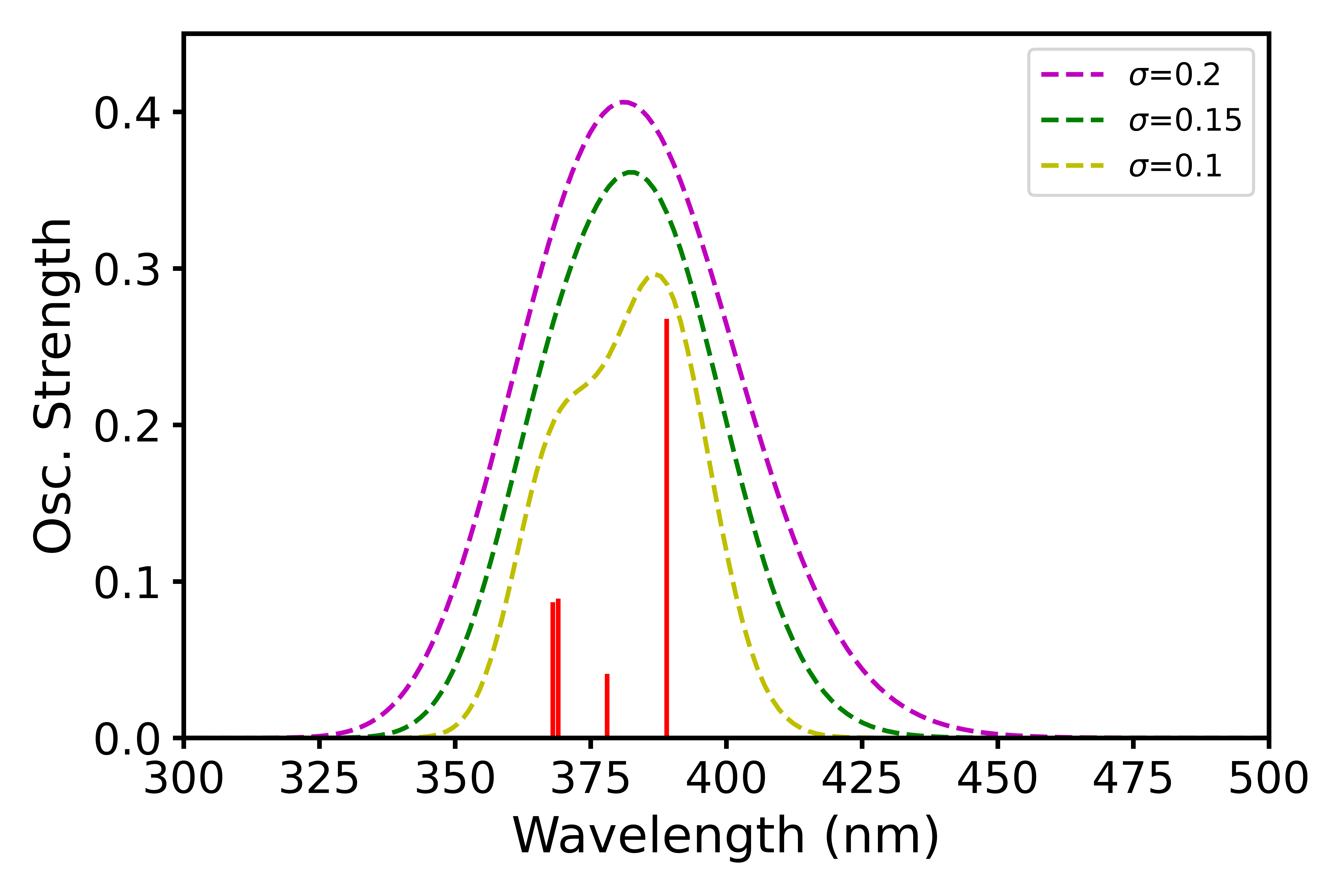}
  \caption{Predicted spectra for 1:1 sums of the \ce{[PoCl5]-} and \ce{[PoCl6]^{2-}} complexes (left) and \ce{[PoCl5(H2O)]-} and \ce{[PoCl6]^{2-}} complexes (right) at the solvated geometries (polarisable continuum model) in the experimental domain of interest (300--500 nm which corresponds to 4.13--2.48 eV in terms of excitation energies). Note that only the lowest-lying absorbing levels are considered, corresponding to or correlating with the first $E$ and $A_1$ levels of \ce{[PoCl5]-} and the first $T_{1u}$ level of \ce{[PoCl6]^{2-}} (see text).}
  \label{fgr:3}
\end{figure*}

By analogy with what came out of EXAFS analysis of bismuth data \cite{Etschmann:2016}, we end up by assessing the possibility for a mixture of \ce{[PoCl5]-} and \ce{[PoCl6]^{2-}} absorbing at 418 nm instead of a pure \ce{[PoCl5]-} or \ce{[PoCl6]^{2-}} species, as in a previous work \cite{stoianov2019uv}. We recall here that in the case of the \ce{Bi^{III}} ion at high HCl concentration, a 1:1 mixture of \ce{[BiCl5]^{2-}} and \ce{[BiCl6]^{3-}} was successfully used to fit the data, but that no sign for the presence of an oxygen atom, which could indicate the coordination of a solvent water molecule, was found. Therefore, in the present work, we have assessed two possibilities, (i) a 1:1 mixture of \ce{[PoCl5]-} and \ce{[PoCl6]^{2-}} at the solvated geometries (\autoref{fgr:3}, left) and a 1:1 mixture of the \ce{[PoCl5(H2O)]-} and \ce{[PoCl6]^{2-}} still at the solvated geometries (\autoref{fgr:3}, right). In practice, since oscillator strengths do not perfectly match the experimental absorbance, we have computed the sum of the spectra of the two species of interest (without dividing it by a factor of 2 to respect the molar fraction), and we have focussed on a domain that is spread around the experimental \SI{418}{\nm} value. 

With the bare \ce{[PoCl5]-} complex at the solvated geometry, one may expect two peaks for small broadening $\sigma$ values, a shoulder on the right of the peak for moderate values, and only a spreading to the right for later values. This is due to the fact that the excitation energies of interest are smaller in the \ce{[PoCl5]-} complex than in the \ce{[PoCl6]^{2-}} one, hence the larger wavelength. When the solvated \ce{[PoCl5(H2O)]-} system is considered, its contribution now appears at a smaller wavelength respective to \ce{[PoCl6]^{2-}}, and actually the difference in wavelength between them is less pronounced. Consequently, if broadening or shouldering there would be, it would clearly sit on the left side of the main peak. Moreover, even a small broadening parameter value such as \SI{0.15}{\electronvolt} points to the potential occurrence of a single peak. Note that the \SI{0.15}{\electronvolt} value has not been randomly chosen, it is the one that eventually leads to a broadening that matches the experimental one ($\sim$\SI{80}{\nm}). Since the experimental peak does not display any shoulder and broadening for larger wavelengths, but rather appears symmetrical, we conclude here that a 1:1 mixture of the solvated \ce{[PoCl5(H2O)]-} and \ce{[PoCl6]^{2-}} systems is also compatible with the experimental data, and not only the pure \ce{[PoCl6]^{2-}} complex, as before concluded\cite{stoianov2019uv}. This mixture hypothesis, which arises from our new quantum mechanical study, would somehow bind up the story in the sense that (i) it is compatible with the data observed in the \ce{Bi^{III}} analogous case and (ii) it would support the conclusion of \citet{bagnall1956544},  complementing previous works which concluded for the occurrence of the sole \ce{[PoCl6]^{2-}} complex \cite{danon1957ion, sheppard1964distribution, marcus1967metal, younes2017solvent, stoianov2019uv}.

\section{Conclusions}

In this work, we have revisited the electronic absorption properties of \ce{Po^{IV}} systems, from the \ce{Po^{4+}} free ion to solvated complexes of experimental relevance in an aqueous solution. 

From a pure electronic structure theory viewpoint, we have shown that the spin-orbit polarisation is very important in the \ce{Po^{4+}} free ion, as it is the case for its analogous \ce{Bi^{3+}} one\cite{real2006ab}. Since the uc-SOCI/SC-NEVPT2 and uc-SOCI/FIC-NEVPT2 approaches lead to satisfactory results in the \ce{Bi^{3+}} case, we believe that our predicted spectra for \ce{Po^{4+}} is of the same quality. In the complexes of interest, we have found small spin-orbit polarisation effects, and that the cost-effective c-SOCI/SC-NEVPT2 approach is particularly sound, which supports previous research\cite{stoianov2019uv}.

From a chemical point of view, we have seen that solvation may be quite tricky in polonium chloride complexes and that a combination of implicit and explicit solvation is required to converge reasonably the predicted excitation energies. In practice, convergence is reached when the first coordination sphere is complete. We recall here that since symmetry matters, partial completion of a given sphere, as the second coordination one, may lead to spurious symmetry lowering, and thus may not lead to better production calculations despite the will for better describing solvation.

Concerning the speciation of \ce{Po^{IV}} in HCl solutions, a lot remains to be done especially from the experimental viewpoint. Our quantum chemical calculations have pointed out that a 1:1 mixture of \ce{[PoCl5(H2O)]-} and \ce{[PoCl6]^{2-}} could very well occur, which naturally remains to be further experimentally probed, for instance by performing EXAFS at high HCl concentration.

Finally, occurrence of a 1:1 mixture of \ce{[PoCl5(H2O)]-} and \ce{[PoCl6]^{2-}} at high HCl concentration (418 nm peak) may also transfer into a another 1:1 mixture at more moderate HCl concentration (corresponding to the 344 nm peak reported by Moyer\cite{moyer1956polonium}), in particular of \ce{[PoCl3(OH)(H2O)2]} and \ce{[PoCl4(OH)(H2O)]^{-}}, which would also have to be later investigated.

\section*{Author Contributions}
N.Z performed all the calculations. R.M. wrote the first draft of the manuscript. All the authors contributed to the scientific development of this work and to the final version of the manuscript.

\section*{Conflicts of interest}
There are no conflicts of interest to declare.

\section*{Acknowledgements}
This work has been supported by the MITI of CNRS (80|Prime project MSM4Po). HPC resources from the CCIPL (``Centre de calcul intensif des Pays de la Loire'') have been used. FR and VV acknowledge support from PIA ANR project CaPPA (ANR-11-LABX-0005-01), I-SITE ULNE projects OVERSEE and MESONM International Associated Laboratory (LAI) (ANR-16-IDEX-0004), and the French Ministry of Higher Education and Research, region Hauts de France council and European Regional Development Fund (ERDF) projects CPER CLIMIBIO and WaveTech.



\balance


\bibliography{References} 
\bibliographystyle{rsc} 

\clearpage



\section*{Supplementary material}

# 1 Molecular structures

All the structures have been obtained at the MP2/def2-TZVP level of theory (see main text for more details).

## 1.1 Gas phase structures

1. [PoCl$_5$]$^-$:

6

Po 0.00000 0.00000 0.00000
Cl 0.00000 2.52599 0.00000
Cl 0.00000 0.00000 2.55781
Cl 2.18757 -1.26300 0.00000
Cl 0.00000 0.00000 -2.55781
Cl -2.18757 -1.26300 0.00000

2. [PoCl$_6$]$^{2-}$:

7

Po 0.00000 0.00000 0.00000
Cl 0.00000 0.00000 2.60400
Cl 0.00000 2.60400 0.00000
Cl 2.60400 0.00000 0.00000
Cl 0.00000 0.00000 -2.60400
Cl -2.60400 0.00000 0.00000
Cl 0.00000 -2.60400 0.00000

3. [PoCl$_5$(H$_2$O)]$^-$:

9

Po -0.06000 -0.00000 0.01300
Cl -1.62200 1.79100 -0.84800
Cl -1.62200 -1.79100 -0.84800
Cl -0.85200 -0.00000 2.37400
Cl 1.81200 1.84600 0.23400
Cl 1.81200 -1.84600 0.23400
O 1.20700 0.00000 -2.09800
H 1.68900 -0.76800 -1.90300
H 1.68900 0.76800 -1.90300

4. [PoCl$_5$(H$_2$O)$_2$]$^-$:

12

Po 0.22448 -0.02103 -0.02614
Cl 1.21980 2.14485 0.91823
O -1.23055 0.39504 1.95667
H -2.17313 0.30679 1.64945
H -1.05685 1.33640 2.11044
O -3.56325 -0.06500 0.71412
H -3.24727 -0.93499 0.41369
H -3.30254 0.50556 -0.03347
Cl -1.30851 -2.09078 -0.40216
Cl 1.63513 -0.22515 -2.05509
Cl -1.59506 1.45125 -1.24523
Cl 1.77060 -1.40296 1.41302

5. [PoCl$_6$(H$_2$O)]$^{2-}$:

10

Po 0.23500 0.00000 -0.00000
Cl 2.05000 1.82600 -0.00000
Cl 0.23000 0.00000 2.59300
O -4.31900 -0.00000 0.00000
Cl -1.62400 -1.87900 0.00000
Cl 0.23000 -0.00000 -2.59300
Cl -1.62400 1.87900 -0.00000
Cl 2.05000 -1.82600 0.00000
H -3.76600 -0.72800 0.00000
H -3.76600 0.72800 -0.00000

**1.2 Solvated structures (polarisable continuum model)**

1. [PoCl$_5$]$^-$:

6

Po 0.00000 0.00000 0.20719
Cl -0.00000 2.55845 0.29454
Cl 0.00000 0.00000 -2.20194
Cl -2.55845 0.00000 0.29454
Cl -0.00000 -2.55845 0.29454
Cl 2.55845 -0.00000 0.29454

2. [PoCl$_6$]$^{2-}$:

7

Po 0.00000 0.00000 0.00000
Cl -0.00000 0.00000 2.58932
Cl -0.00000 2.58932 -0.00000
Cl 2.58932 0.00000 0.00000
Cl -0.00000 -0.00000 -2.58932
Cl -2.58932 0.00000 -0.00000
Cl -0.00000 -2.58932 0.00000

3. [PoCl$_5$(H$_2$O)]$^-$:

9

Po -0.00082 -0.00000 -0.03248
Cl -2.57448 -0.00027 -0.11770
Cl 0.00269 -2.55349 -0.33109
Cl -0.02789 0.00005 2.47370
Cl 0.00216 2.55347 -0.33116
Cl 2.57505 0.00026 -0.06290
O 0.02637 -0.00002 -2.40170
H 0.11994 -0.78585 -2.89143
H 0.11986 0.78584 -2.89141

4. [PoCl$_5$(H$_2$O)$_2$]$^-$:

12

Po -0.20000 -0.00800 0.00900
Cl -0.95800 2.29200 -0.86000
O 1.34400 0.05600 -1.86400
H 2.29600 0.03300 -1.54800
H 1.24200 0.86400 -2.39200
O 3.73500 0.05400 -0.78400
H 4.19200 -0.79400 -0.69800
H 3.48800 0.30800 0.12100
Cl 0.89500 -2.27100 0.52400
Cl -1.78900 -0.06600 1.93500
Cl 1.64800 1.15500 1.43300
Cl -1.85900 -1.14700 -1.56200

5. [PoCl$_6$(H$_2$O)]$^{2-}$:

10

Po -0.22200 -0.00000 -0.00400
Cl -1.90300 1.81700 -0.68200
Cl 0.76400 0.00000 -2.39000
Cl 1.48200 -1.86400 0.68800
Cl -1.17400 -0.00000 2.39700
Cl 1.48200 1.86400 0.68900
Cl -1.90300 -1.81700 -0.68300
O 4.09600 -0.00000 -0.02600
H 3.58500 -0.74500 0.11700
H 3.58500 0.74400 0.11700

## 2 Supplementary tables

Table S1 Numerical compositions the excited energy levels expressed in terms of the spin-orbit free states, for the bare [PoCl$_5$]$^-$ and [PoCl$_6$]$^{2-}$ complexes in the gas phase, obtained at the c-SOCI/SC-NEVPT2 level.

| System | SPG | Symmetry | Composition |
|---|---|---|---|
| [PoCl$_5$]$^-$ | $D_{3h}$ | $A_1''$ | 81.3% *$^3E'$ + 18.7% *$^3A_2''$ |
| | | $A_2''$ | 86.9% *$^3E'$ + 13.1% *$^1A_2''$ |
| | | $E'$ | 48.2% *$^3E'$ + 20.8% *$^3A_2''$ + 31.0% *$^1E'$ |
| | | $E''$ | $^3E'$ |
| | | $E'$ | 46.9%' *$^3E'$ + 3.3%' *$^3A_2''$ + 49.8% *$^1E'$ |
| | | $A_1''$ | 18.7% *$^3E'$ + 81.3% *$^3A_2''$ |
| | | $E'$ | 4.9% *$^3E'$ + 75.9% *$^3A_2''$ + 19.2% *$^1E'$ |
| | | $A_2''$ | 13.1% *$^3E'$ + 86.9% *$^1A_2''$ |
| [PoCl$_6$]$^{2-}$ | $O_h$ | $A_{1u}$ | $^3T_{1u}$ |
| | | $T_{1u}$ | 73.8% *$^3T_{1u}$ + 26.2% *$^1T_{1u}$ |
| | | $T_{2u} \oplus E_u$ | $^3T_{1u}$ |
| | | $T_{1u}$ | 26.2% *$^3T_{1u}$ + 73.8% *$^1T_{1u}$ |

Table S2 Numerical compositions the excited energy levels expressed in terms of the spin-orbit free states, for the bare [PoCl$_5$]$^-$ and [PoCl$_6$]$^{2-}$ complexes at the solvated geometries (polarisable continuum model), obtained at the c-SOCI/SC-NEVPT2 level.

| System | SPG | Symmetry | Composition |
|---|---|---|---|
| [PoCl$_5$]$^-$ | $C_{4v}$ | $A_2$ | 68.6% $^3A_1$ + 31.3% *$^3E$ |
| | | $E$ | 74.8% *$^3A_1$ + 15.5% *$^3E$ + 9.6% *$^1E$ |
| | | $A_1$ | 36.3% *$^3E$ + 63.1% *$^1A_1$ |
| | | $A_2$ | 31.3% *$^3A_1$ + 68.7% *$^3E$ |
| | | $A_1$ | 62.9% *$^3E$ + 36.7% *$^1A_1$ |
| | | $E$ | 24.4% *$^3A_1$ + 59.3% *$^3E$ + 16.3% *$^1E$ |
| | | $B_1 \oplus B_2$ | $^3E$ |
| | | $E$ | 0.6% *$^3A_1$ + 25.2% *$^3E$ + 74.0% *$^1E$ |
| [PoCl$_6$]$^{2-}$ | $O_h$ | $A_{1u}$ | $^3T_{1u}$ |
| | | $T_{1u}$ | 73.9% *$^3T_{1u}$ + 26.1% *$^1T_{1u}$ |
| | | $T_{2u} \oplus E_u$ | $^3T_{1u}$ |
| | | $T_{1u}$ | 26.1% *$^3T_{1u}$ + 73.9% *$^1T_{1u}$ |

Table S3 Spin character of the excited energy levels and excitation energies (in eV), for the [PoCl$_5$(H$_2$O)]$^-$ complex at the solvated geometry (polarisable continuum model), obtained at the c-SOCI/SC-NEVPT2 level.

| System | SPG | Composition | Energy |
|---|---|---|---|
| [PoCl$_5$(H$_2$O)]$^-$ | $C_1$ | triplet | 3.21 |
| | | triplet/singlet | 3.28 |
| | | triplet/singlet | 3.36 |
| | | triplet/singlet | 3.37 |
| | | triplet | 5.14 |
| | | triplet/singlet | 5.17 |
| | | triplet/singlet | 5.20 |
| | | triplet/singlet | 5.32 |
| | | triplet | 5.40 |
| | | triplet/singlet[a] | 5.41 |
| | | triplet/singlet | 5.70 |
| | | triplet/singlet | 5.84 |

[a] This energy level correlates with a pure spin-triplet energy level in the reference $C_{4v}$ structure of the [PoCl$_5$]$^-$ complex. Because of the symmetry lowering to $C_1$, some little spin-singlet character has emerged. This is not the case of the energy levels that are here indicated as "triplet" (for which no spin-singlet character was numerically observed).



\end{document}